\documentclass[twocolumn,nofootinbib,prl]{revtex4}

\usepackage{graphicx}
\usepackage{xcolor}
\usepackage[hypertex]{hyperref}
\usepackage{amsmath, amssymb}

\newcommand{\lsim}{\lesssim}
\newcommand{\ord}[1]{\mathcal{O}{(#1)}}

\newcommand{\gsim}{\gtrsim}

\newcommand{\beq}{\begin{equation}}
\newcommand{\eeq}{\end{equation}}
\newcommand{\bea}{\begin{eqnarray}}
\newcommand{\eea}{\end{eqnarray}}

\newcommand{\eps}{\varepsilon}

\newcommand{\tev}{{\rm TeV}}
\newcommand{\gev}{{\rm GeV}}
\newcommand{\mev}{{\rm MeV}}

\begin{document}
\pagestyle{plain}
\title{Muon Anomaly and Dark Parity Violation}
\author{Hooman Davoudiasl\footnote{email: hooman@bnl.gov}}
\author{Hye-Sung Lee\footnote{email: hlee@bnl.gov}}
\author{William J. Marciano\footnote{email: marciano@bnl.gov}}
\affiliation{Department of Physics, Brookhaven National Laboratory, Upton, NY 11973, USA}
\date{May 2012}

\begin{abstract}
The muon anomalous magnetic moment exhibits a $3.6 \sigma$ discrepancy between experiment and theory.
One explanation requires the existence of a light vector boson, $Z_d$ (the dark $Z$), with mass $10 - 500 ~\mev$ that couples weakly to the electromagnetic current through kinetic mixing.
Support for such a solution also comes from astrophysics conjectures regarding the utility of a $U(1)_d$ gauge symmetry in the dark matter sector.
In that scenario, we show that mass mixing between the $Z_d$ and ordinary $Z$ boson introduces a new source of ``dark'' parity violation which is potentially observable in atomic and polarized electron scattering experiments.
Restrictive bounds on the mixing $(m_{Z_d} / m_Z) \delta$ are found from existing atomic parity violation results, $\delta^2 < 2 \times 10^{-5}$.
Combined with future planned and proposed polarized electron scattering experiments, a sensitivity of 
$\delta^2 \sim 10^{-6}$ is expected to be reached, thereby complementing direct searches for the $Z_d$ boson.
\end{abstract}
\maketitle

For a number of years, there has been a persistent disagreement between the experimental value of the muon anomalous magnetic moment, $a_\mu \equiv (g_\mu-2)/2$
\beq
a_\mu^{\rm exp} = 116\ 592\ 089 (63) \times 10^{-11}
\eeq
and the theoretical $SU(3)_C \times SU(2)_L \times U(1)_Y$ Standard Model (SM) prediction
\beq
a_\mu^{\rm SM} = 116\ 591\ 802 (49) \times 10^{-11} .
\eeq
The above $3.6 \sigma$ discrepancy \cite{PDG}
\beq
\Delta a_\mu = a_\mu^{\rm exp} - a_\mu^{\rm SM} = 287 (80) \times 10^{-11} \label{eq:3}
\eeq
could be indicative of problems with the theoretical calculations and/or experimental measurements.
Alternatively, it could be a harbinger of ``new physics'' effects beyond SM expectations \cite{Czarnecki:2001pv}.
One possibility, receiving support from dark matter conjectures 
\cite{DMzprime,Davoudiasl:2010am}, envisions the existence of a 
relatively light $U(1)_d$ gauge boson, $Z_d$, coming from the ``dark'' 
sector that indirectly couples to our world via $U(1)_Y \times U(1)_d$ 
kinetic mixing \cite{Holdom:1985ag}, parametrized by $\eps$ such that \cite{Davoudiasl:2012ag}
\beq
{\cal L}_\text{int} = -e \eps Z_d^\mu J_\mu^{em}, \quad J_\mu^{em}         = \sum_f Q_f \bar f \gamma_\mu f ,
\label{eq:4}
\eeq
where $Q_f$ is the electric charge of fermion $f$.
The coupling of $Z_d$ to the weak neutral current from kinetic mixing is suppressed at low energies because of a cancellation between the $\eps$ dependent field redefinition and leading $Z$-$Z_d$ mass matrix diagonalization effects induced by $\eps$ \cite{Davoudiasl:2012ag}.
(We do not consider here the possibility that some ordinary fermions may have explicit $U(1)_d$ charges.)

The $Z_d \mu \bar\mu$ vector current coupling in Eq.~\eqref{eq:4} gives rise to an additional one loop contribution \cite{Fayet:2007ua,Leveille:1977rc} to $a_\mu$
\bea
&& a_\mu^{Z_d} (\text{vector}) = \frac{\alpha}{2\pi} \eps^2 F_V\left(m_{Z_d} / m_\mu \right) \label{eq:5} \\
&& F_V(x) \equiv \int_0^1 dz \frac{2 z (1-z)^2}{(1-z)^2 + x^2 z} , \quad F_V(0) = 1 .  \label{eq:6}
\eea

\begin{figure}[t]
\begin{center}
\includegraphics[width=0.39\textwidth]{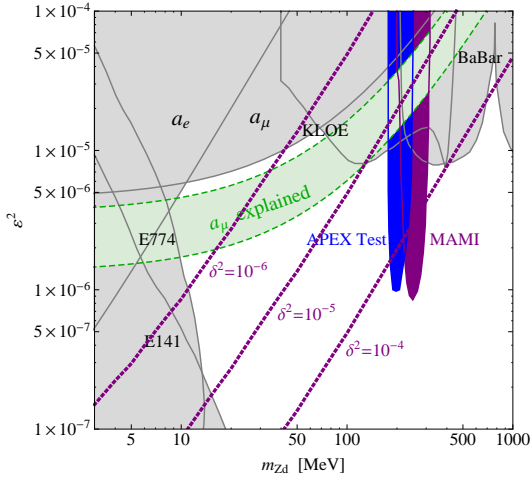}
\end{center}
\caption{Dark $Z$ boson exclusion regions (partly adapted from Ref.~\cite{McKeown:2011yj}) in the $(m_{Z_d}, \eps^2)$ plane along with the band that explains the $\Delta a_\mu$ discrepancy ($90 \%$ CL) and exclusion regions from atomic parity violation (above the lines) for $Z$-$Z_d$ mixing $\delta$ values.}
\label{fig:1}
\end{figure}

The effect in Eq.~\eqref{eq:5} has the right algebraic sign, such that for $10 ~\mev \lsim m_{Z_d} \lsim 500 ~\mev$ and $\eps^2$ roughly in the range $10^{-6} - 10^{-4}$, the discrepancy $\Delta a_\mu$ in Eq.~\eqref{eq:3} can be eliminated.
We plot \cite{McKeown:2011yj} in Fig.~\ref{fig:1} the band in $(m_{Z_d}, \eps^2)$ space that reduces the discrepancy to within $90\%$ CL, {\it i.e.}
\beq
a_\mu^{Z_d} = 287 \pm 131 \times 10^{-11} .
\eeq
There, we also give a (roughly) $90\%$ CL bound from the electron anomalous magnetic moment \cite{Czarnecki:1900zz,Hanneke:2008tm} constraint $|a_e^{Z_d}| < 10^{-11}$ using $m_e$ in place of $m_\mu$ in Eq.~\eqref{eq:5} as well as a $3\sigma$ $a_\mu^{Z_d}$ bound.
Constraints from other direct experimental searches for $Z_d$ are also given \cite{Bjorken:2009mm,Abrahamyan:2011gv}.
However, those bounds are somewhat model dependent since they assume the $Z_d$ decays primarily into $e^+e^-$ or $\mu^+\mu^-$ pairs.
They will be diluted if, for example, $Z_d$ decays primarily into light ``dark particles'' that escape the detector as $Z_d \to$ missing energy \cite{Davoudiasl:2012ag}.

Recently \cite{Davoudiasl:2012ag}, we generalized the $U(1)_d$ kinetic mixing scenario to include possible $Z$-$Z_d$ mass mixing by introducing the $2\times2$ mass matrix
\beq
M_0^2 = \begin{pmatrix}1 & -\eps_Z \\ 
-\eps_Z  & ~~m_{Z_d}^2/m_Z^2\end{pmatrix} m_Z^2 
\label{eq:9}
\eeq
where $m_{Z_d}$ and $m_Z$ (with $m_{Z_d}^2 \ll m_Z^2$) represent the ``dark'' $Z$ and SM $Z$ masses (before diagonalization).
The off-diagonal mixing is parametrized by
\beq
\eps_Z = \frac{m_{Z_d}}{m_Z} \delta , \quad 0 \le |\delta| < 1
\eeq
where the $m_{Z_d}/m_Z$  factor allows a smooth $m_{Z_d} \to 0$ limit for nonconserved current amplitudes and $\delta$ is expected to be a small quantity that depends on the Higgs scalar sector of the theory \cite{Davoudiasl:2012ag}.
$Z$-$Z_d$ mixing induced by $\eps_Z$ leads to an additional coupling of $Z_d$ to fermions via the weak neutral current
\bea
\begin{split}
{\cal L}_\text{int} &= -\frac{g}{2 \cos\theta_W} \eps_Z Z_d^\mu J_\mu^{NC} \\
J_\mu^{NC} &= \sum_f (T_{3f} - 2 Q_f \sin^2\theta_W) \bar f \gamma_\mu f - T_{3f} \bar f \gamma_\mu \gamma_5 f
\end{split}
\label{eq:11}
\eea
with $T_{3f} = \pm 1/2$ and $\sin^2\theta_W \simeq 0.23$ the SM weak mixing angle.
Because of its axial-vector coupling, this new interaction violates parity and current conservation.
As a result, it can lead to potentially observable effects in atomic parity violation (APV) and polarized electron scattering experiments, as well as rare flavor changing $K$ and $B$ or Higgs boson decays ($H \to Z Z_d$) to longitudinally polarized $Z_d$ bosons (phase space permitting).
We pointed out in Ref.~\cite{Davoudiasl:2012ag} that the nonobservation of such effects already leads to bounds $|\delta| \lsim 10^{-2} - 10^{-3}$ depending on $m_{Z_d}$ and in some cases $\eps$.
Here, we further explore such constraints, but focus on that part of parameter space $10 ~\mev \lsim m_{Z_d} \lsim 500 ~\mev$ and $|\eps| \approx 10^{-3} - 10^{-2}$ favored by a $Z_d$ explanation of the $\Delta a_\mu$ discrepancy in Eq.~\eqref{eq:3}.
Also, to keep our analysis independent of the $Z_d$ decay properties, we concentrate on low energy parity violation, {\it i.e.}  atomic and polarized electron scattering experiments.
A variety of direct searches for $Z_d$ have been discussed in the literature 
\cite{Davoudiasl:2012ag,McKeown:2011yj,Bjorken:2009mm,Abrahamyan:2011gv}.

\begin{figure*}[t]
\begin{center}
\includegraphics[width=0.39\textwidth]{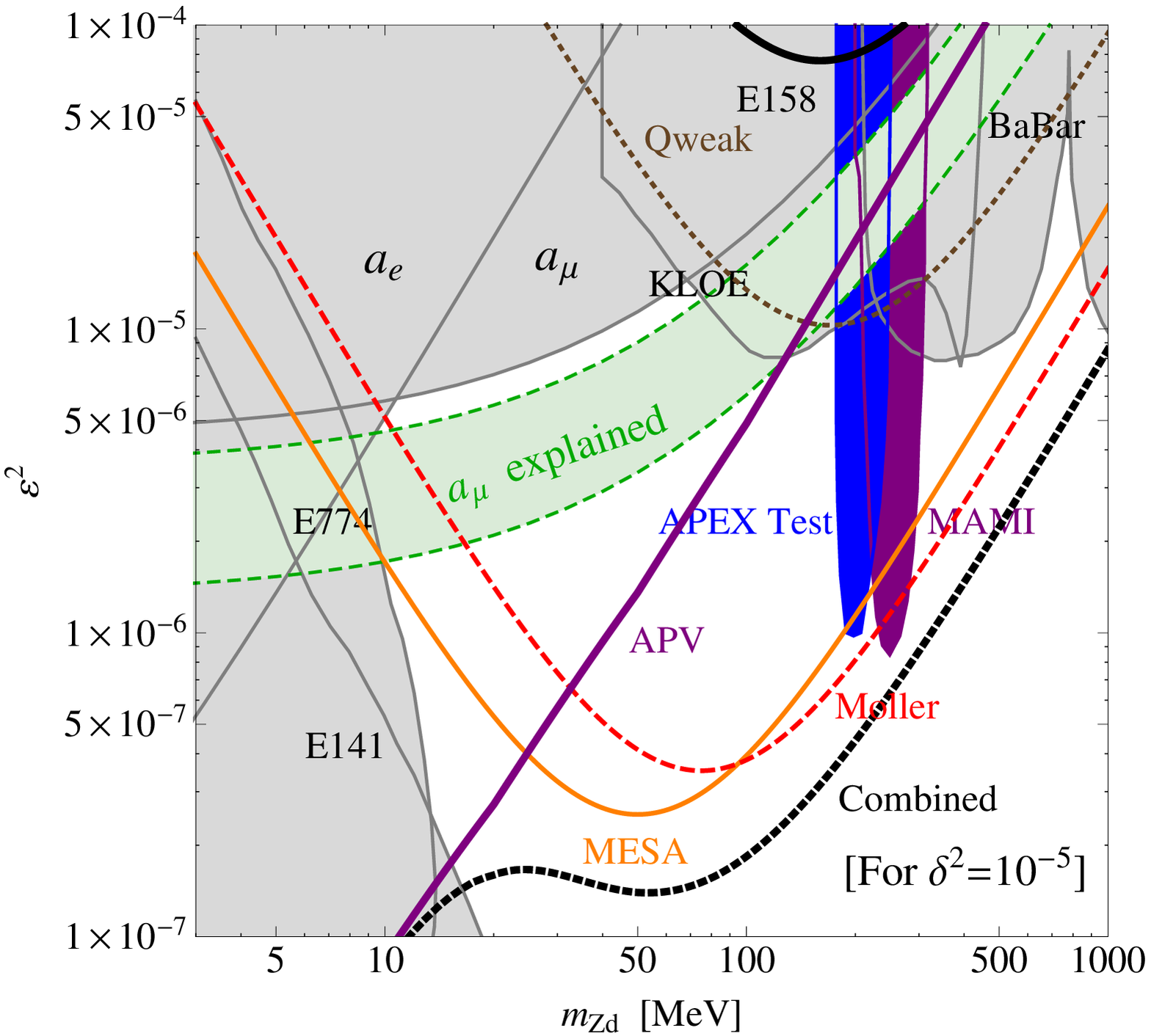} ~~~
\includegraphics[width=0.39\textwidth]{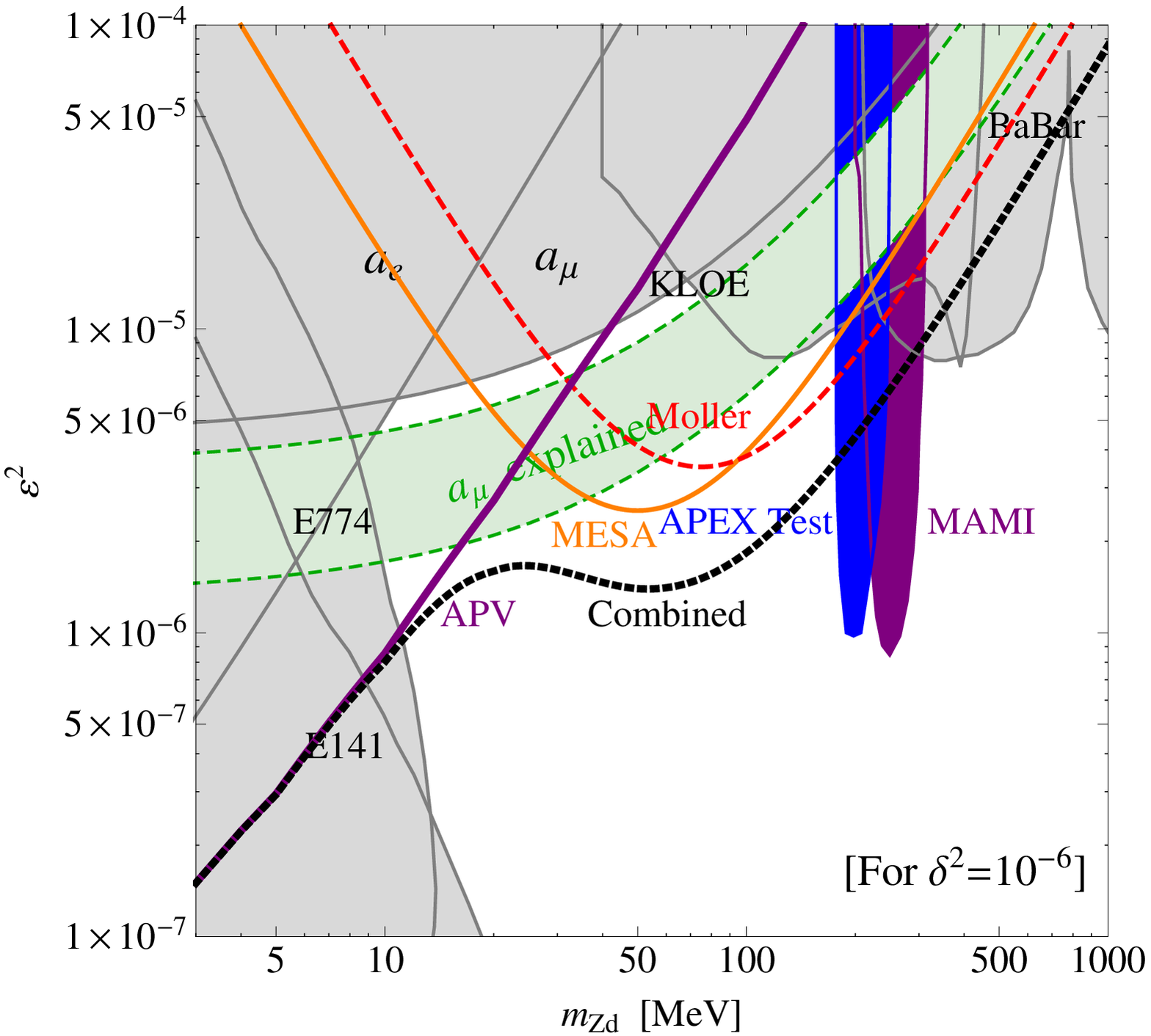}
\end{center}
\caption{Dark $Z$ boson exclusion regions from various parity violating experiments (existing and proposed) and their combined sensitivity for $\delta^2 = 10^{-5}$ (Left) and $10^{-6}$ (Right) at $90 \%$ CL.}
\label{fig:2}
\end{figure*}

We begin by considering changes to $a_\mu^{Z_d}$ due to $\delta \ne 0$.
The additional $Z_d \mu \bar\mu$ vector coupling in Eq.~\eqref{eq:11} modifies the contribution in Eq.~\eqref{eq:6} via the replacement
\beq
\eps^2 \to \left( \eps + \eps_Z \frac{1 - 4 \sin^2\theta_W}{4 \sin\theta_W \cos\theta_W} \right)^2 \simeq (\eps + 0.02 \eps_Z)^2
\label{eps2shift}
\eeq
where $\sin^2\theta_W \simeq 0.24$ appropriate for low $Q^2 \simeq m_\mu^2$ scales \cite{Czarnecki:1995fw} has been employed.
For the $\Delta a_\mu$ favored range of $m_{Z_d}$ and $\eps^2$ in Fig.~\ref{fig:1}, the shift 
in Eq.~(\ref{eps2shift}) is small ($\lsim 2 \%$) for all $\delta$ and can be ignored.

The axial-vector part of the $Z_d \mu \bar\mu$ coupling in Eq.~\eqref{eq:11} gives rise to a negative contribution \cite{Leveille:1977rc}
\bea
\begin{split}
a_\mu^{Z_d} (\text{axial}) &= -\frac{G_F m_\mu^2}{8 \sqrt{2} \pi^2} \delta^2 F_A \left( m_{Z_d} / m_\mu \right) \\
&\simeq - 117 \times 10^{-11} \delta^2 F_A \left( m_{Z_d}/m_\mu \right) 
\end{split} \\
F_A(x) \equiv \int_0^1 dz \frac{2 (1-z)^3 + x^2 z (1-z) (z+3)}{(1-z)^2 + x^2 z},
\eea
where $G_F \simeq 1.166 \times 10^{-5} ~\gev^{-2}$, $F_A(0) = 1$ , and $F_A(\infty) = 5/3$.  For $\delta^2 \lsim 0.1$ (a mild requirement \cite{Davoudiasl:2012ag}), that contribution is also negligible throughout the $\Delta a_\mu$ favored region in Fig.~\ref{fig:1}.
So, we conclude that the effect of $Z$-$Z_d$ mass mixing plays little direct role in any discussion of the $\Delta a_\mu$ discrepancy and its interpretation as due to $\eps^2$.

Next, we examine constraints on the $m_{Z_d}$, $\eps$, $\delta$ parameter space coming from low energy parity violating experiments and their implications for a $Z_d$ interpretation of the $\Delta a_\mu$ discrepancy.

It is well known that the classic Cesium atomic parity violation experiment \cite{Bennett:1999pd} provides a stringent constraint on heavy $Z'$ bosons \cite{Marciano:1990dp} that violate parity, often implying $m_{Z'} \gsim \ord{1 ~\tev}$.
However, its application to relatively light gauge bosons such as $Z_d$ has been less explored.
Such a connection was first made by Bouchiat and Fayet \cite{Bouchiat:2004sp} for a light ``$U$-boson'' with very general parity violating couplings to fermions.
They found strong constraints and argued against axial-vector couplings.
We recently \cite{Davoudiasl:2012ag} revisited the application of low energy parity violation experimental constraints within the general $Z$-$Z_d$ mass mixing  formalism of Eq.~\eqref{eq:9}.
We updated the Cesium constraint to include more recent atomic theory \cite{Porsev:2009pr}, expanded the analysis to polarized electron scattering \cite{Derman:1979zc} and applied our study specifically to the ``dark'' $Z$ boson.
Here, we focus on the connection of that analysis with the $\Delta a_\mu$ discrepancy and its interpretation via $10 ~\mev \lsim m_{Z_d} \lsim 500 ~\mev$ with $\eps^2 \sim 10^{-6} - 10^{-4}$.

The additional parity violation from Eq.~\eqref{eq:11} manifests itself as replacements in low energy SM parity violating weak neutral current amplitudes \cite{Davoudiasl:2012ag}
\beq
G_F            \to \rho_d G_F , \quad \sin^2\theta_W \to \kappa_d \sin^2\theta_W ,
\eeq
where for (momentum transfer) $Q^2 = - q^2$
\bea
&& \rho_d = 1 + \delta^2 f\left( Q^2 / m_{Z_d}^2 \right) , \\
&& \kappa_d = 1 - \eps \delta \frac{m_Z}{m_{Z_d}} \frac{\cos\theta_W}{\sin\theta_W} f\left( Q^2 / m_{Z_d}^2 \right)
\eea
giving rise to
\beq
\Delta \sin^2\theta_W \simeq - 0.42 \eps \delta \frac{m_Z}{m_{Z_d}} f\left(Q^2 / m_{Z_d}^2 \right) .
\label{eq:new17}
\eeq
As pointed out in Ref.~\cite{Bouchiat:2004sp}, for parity violation in heavy atoms, such as Cesium, 
there is a correction factor $f = K(\rm{Cs})$ relevant for very small $m_{Z_d}$.
For example, $K(\rm{Cs})\simeq 0.5$ at $m_{Z_d} \simeq 2.4 ~\mev$, which sets the typical momentum transfer $\left< Q \right>$ in this case, whereas $K(\rm{Cs}) \simeq 0.74$, $0.98$ at $m_{Z_d} \simeq 10$, $100 ~\mev$.
In the case of polarized electron scattering asymmetries, the $Z_d$ propagator effect gives
\beq
f\left( Q^2 / m_{Z_d}^2 \right) = \frac{1}{1 + Q^2 / m_{Z_d}^2}
\eeq
with $\left< Q \right>$ ranging from $50 - 170 ~\mev$ for the experiments we consider.

Currently, the SM prediction for the weak nuclear charge $Q_W (Z,N) \simeq -N + Z(1 - 4 \sin^2\theta_W)$ in the case of $^{133}_{55} {\rm Cs}$ (including electroweak radiative corrections) \cite{Marciano:1982mm}
\beq
Q_W^\text{SM} (^{133}_{55} {\rm Cs}) = -73.16(5)
\eeq
is in excellent agreement with experiment (including the most up-to-date atomic theory) \cite{Bennett:1999pd,Porsev:2009pr}
\beq
Q_W^\text{exp} (^{133}_{55} {\rm Cs}) = -73.16(35) .
\eeq
The $90 \%$ CL bound on the difference
\beq
| \Delta Q_W ({\rm Cs}) | = | Q_W^\text{exp} (^{133}_{55} {\rm Cs}) - Q_W^\text{SM} (^{133}_{55} {\rm Cs}) | < 0.6
\label{eq:26}
\eeq
can be compared with the potential $Z_d$ contribution \cite{Davoudiasl:2012ag}
\beq
\begin{split}
&\Delta Q_W (^{133}_{55} {\rm Cs}) = \\
&\left( -73.16 \delta^2 + 220 \eps \delta \frac{m_Z}{m_{Z_d}} \sin\theta_W \cos\theta_W \right) K({\rm Cs}) .
\end{split}
\label{eq:27}
\eeq

\begin{table*}[tb]
\begin{tabular}{|c|r|l|l|}
\hline
~~Experiment~~   & ~$\left<Q\right>$~~~~~ & ~$\sin^2\theta_W$($m_Z$)~ & ~~~~Bound on dark $Z$ ~($90\%$ CL)~~ \\
\hline
~Cesium APV~     & ~$2.4 ~\mev$~ & ~~$0.2313(16)$~  & ~$\eps^2 < \frac{39 \times 10^{-6}}{\delta^2} \left( \frac{m_{Z_d}}{m_Z} \right)^2 \frac{1}{K(m_{Z_d})^2}$~ \\
~E158 (SLAC)~    & ~$160 ~\mev$~ & ~~$0.2329(13)$~  & ~$\eps^2 < \frac{62 \times 10^{-6}}{\delta^2} \left( \frac{(160 ~\mev)^2 + m_{Z_d}^2}{m_Z \, m_{Z_d}} \right)^2$~ \\
~Qweak (JLAB)~   & ~$170 ~\mev$~ & ~~$\pm 0.0007$~  & ~$\eps^2 < \frac{7.4 \times 10^{-6}}{\delta^2} \left( \frac{(170 ~\mev)^2 + m_{Z_d}^2}{m_Z \, m_{Z_d}} \right)^2$~ \\
~Moller (JLAB)~  & ~$75 ~\mev$~  & ~~$\pm 0.00029$~ & ~$\eps^2 < \frac{1.3 \times 10^{-6}}{\delta^2} \left( \frac{(75 ~\mev)^2 + m_{Z_d}^2}{m_Z \, m_{Z_d}} \right)^2$~ \\
~MESA (Mainz)~   & ~$50 ~\mev$~  & ~~$\pm 0.00037$~ & ~$\eps^2 < \frac{2.1 \times 10^{-6}}{\delta^2} \left( \frac{(50 ~\mev)^2 + m_{Z_d}^2}{m_Z \, m_{Z_d}} \right)^2$~ \\
\hline
~Combined~       &               & ~~$\pm 0.00021$~ & ~$\eps_\text{comb}^2 < \frac{1}{\sum_i (1 / \eps_i^2)}$~ \\
\hline
\end{tabular}
\caption{Existing and possible future constraints on dark $Z$ from various parity violating experiments}
\label{tab:1}
\end{table*}

In principle, there could be a cancellation between the two terms in Eq.~\eqref{eq:27} for $\eps (m_Z / m_{Z_d}) \sim 0.8 \delta$.
However, for the $\Delta a_\mu$ preferred band in Fig.~\ref{fig:1}, $| \eps (m_Z / m_{Z_d}) | \gsim 2$; the second term in Eq.~\eqref{eq:27} always dominates.
In fact, a conservative self-consistent assessment of the bound (at $90 \%$ CL) from Eqs.~\eqref{eq:26} and \eqref{eq:27}
yields
\beq
| \delta^2 - 2 \delta | < 0.008 ~~\to~~ \delta^2 < 2 \times 10^{-5}
\eeq
for the entire $\Delta a_\mu$ motivated band in Fig.~\ref{fig:1}.
That means the first term in Eq.~\eqref{eq:27} can be neglected and the $Q_W (^{133}_{55} {\rm Cs})$ bound becomes for arbitrary $\eps^2$ and $m_{Z_d}$ essentially a bound 
\beq
\eps^2 < \frac{4 \times 10^{-5}}{\delta^2 K^2} \left( \frac{m_{Z_d}}{m_Z} \right)^2 
\eeq
on the allowed $\sin^2\theta_W$ shift.  The atomic parity violation bound on $\eps^2$ is illustrated in Fig.~\ref{fig:1} for various values of $\delta^2$.
Note that for $\delta^2 \gsim 2 \times 10^{-5}$, the entire $\Delta a_\mu$ discrepancy motivated band is already ruled out.
Alternatively, if a light $Z_d$ is responsible for the $\Delta a_\mu$ discrepancy, the $Z$-$Z_d$ mixing $| \eps_Z | = | (m_{Z_d} / m_Z) \delta |$ must be very tiny ($\delta^2 < 2 \times 10^{-5}$).
Of course, the $\Delta a_\mu$ discrepancy may have nothing to do with $Z_d$.
In that case, larger $\delta^2$ values can be accommodated by going to smaller $\eps^2$ or larger $m_{Z_d}$ values, although other constraints \cite{Davoudiasl:2012ag} then come into play.

Atomic parity violation already provides a powerful constraint on $\delta^2$ over an interesting $m_{Z_d}$ range.
Future experiments employing ratios of isotopes could, in principle, eliminate the atomic theory uncertainty and further probe $Z_d$ mass and mixing as well as other ``new physics'' scenarios \cite{Dzuba:1985gw}.

Another type of low energy parity violating experiment involves 
polarized electron scattering on electrons, protons or other targets.
They measure the parity violating asymmetry \cite{Derman:1979zc} $A_{LR} \equiv \sigma_L - \sigma_R / \sigma_L + \sigma_R$ due to $\gamma$-$Z$ interference at low $Q^2$.
In some cases, such as $ee$ and $ep$, those experiments are particularly sensitive to $\sin^2\theta_W$ at low $Q^2$, where the effective $\sin^2\theta_W$ is expected \cite{Czarnecki:1995fw} to be about $0.24$, thereby leading to very small asymmetries (proportional to $1 - 4 \sin^2\theta_W$).
Already, experiment E158 at SLAC has measured \cite{Anthony:2005pm} (evolving to $Q^2 = m_Z^2$)
\beq
\sin^2\theta_W (m_Z)_{\overline{\rm MS}} = 0.2329(13) \quad (\text{E158 at SLAC})
\label{eq:32}
\eeq
which is to be compared with the $Z$ pole average \cite{PDG}
\beq
\sin^2\theta_W (m_Z)_{\overline{\rm MS}} = 0.23125(16) .
\label{eq:33}
\eeq
The relatively good agreement between Eqs.~\eqref{eq:32} and \eqref{eq:33} already constrains many types of ``new physics'' at a sensitivity similar to APV.
In the case of $Z_d$ at low masses, Cesium APV has the advantage of a low \cite{Bouchiat:2004sp} $\left< Q \right> \simeq 2.4 ~\mev$ while for E158, $\left< Q \right>^\text{E158} \simeq 160 ~\mev$ such that $Z_d$ propagator effects suppress the sensitivity by $m_{Z_d}^2 / (Q^2 + m_{Z_d}^2)$ at the amplitude level.

A comparison of E158 constraints, using (see Eq.~\eqref{eq:new17})
\beq
\eps^2 < \frac{6 \times 10^{-5}}{\delta^2} \left( \frac{0.026 ~\gev^2 + m_{Z_d}^2}{m_Z m_{Z_d}} \right)^2
\eeq
with APV, is illustrated in Fig.~\ref{fig:2}.
The one-sided $90 \%$ CL coefficient in that bound has been increased due to the $\sim 1\sigma$ difference between Eqs.~\eqref{eq:32} and \eqref{eq:33}.
For a given $\delta^2$, the bounds at large $m_{Z_d}$ are similar, but APV is superior for $m_{Z_d} \lsim 160 ~\mev$.

An ongoing polarized $ep$ experiment \cite{E-08-016,McKeown:2011yj}, Qweak at JLAB, aims to measure $\sin^2\theta_W$ to $\pm 0.0007$ at $\left< Q \right> \simeq 170 ~\mev$.
That represents an improvement by about a factor of 2 over E158, but the similar $\left< Q \right>$ means that it also lacks low $m_{Z_d}$ sensitivity.
In the longer term, a new polarized $ee$ (Moller) \cite{E-12-09-005} experiment at JLAB would measure $\sin^2\theta_W$ to $\pm 0.00029$ at $\left< Q \right> \simeq 75 ~\mev$, and a very low energy polarized $ep$ experiment at a new proposed MESA facility \cite{MESA} in Mainz, Germany, would measure $\sin^2\theta_W$ to $\pm 0.00037$ for $\left< Q \right>$ perhaps as low as $50 ~\mev$.
The sensitivities of these (proposed) experiments are also illustrated in Fig.~\ref{fig:2}, using 
the constraints in Table~\ref{tab:1} derived from Eq.~\eqref{eq:new17}.

In Fig.~\ref{fig:2}, we give a combined sensitivity bound for $\delta^2 = 10^{-5}$ and $\delta^2 = 10^{-6}$ from all existing and proposed low energy parity violating experiments.
That plot illustrates the complementarity of atomic and polarized electron scattering experiments.
In addition to providing overlapping probes of new physics, collectively they span a large range of ($m_{Z_d}$, $\eps^2$) space and probe down to $\delta^2$ of $\ord{10^{-6}}$.
Of course, it is possible that a light $Z_d$ exists that is consistent with the $\Delta a_\mu$ discrepancy and will be discovered.
For example, if $m_{Z_d} \simeq 75 ~\mev$, $|\eps| \simeq 3 \times 10^{-3}$ and $|\delta| \simeq 2 \times 10^{-3}$, the proposed Moller and MESA experiments should find shifts $| \Delta \sin^2 \theta_W | \simeq 0.0015$ and $0.0021$, respectively, corresponding to about $5 \sigma$ discovery sensitivities.

In conclusion, we have found that existing atomic parity violating results already require $\delta^2 \lsim 2 \times 10^{-5}$ for the entire range of ($m_{Z_d}$, $\eps^2$), {\it i.e.} $10 ~\mev \lsim m_{Z_d} \lsim 500 ~\mev$, $\eps^2 \simeq 10^{-6} - 10^{-4}$, favored by the $Z_d$ interpretation of the $\Delta a_\mu$ discrepancy.
That requirement calls into question the $Z_d$ interpretation of the $\Delta a_\mu$ unless $Z$-$Z_d$ mixing is naturally small, for example, if the mass $m_{Z_d}$ is primarily generated by an $SU(2)_L \times U(1)_Y$ Higgs singlet \cite{Davoudiasl:2012ag}.
Future polarized electron scattering experiments will provide additional $Z_d$ sensitivity, particularly for $m_{Z_d} \gsim 75 ~\mev$ (where $5 \sigma$ effects are possible) and will nicely complement atomic parity violation experiments as well as direct $Z_d$ searches.

\acknowledgments
Acknowledgments: This work was supported in part by the United States Department of Energy under Grant Contract No. DE-AC02-98CH10886.
WM acknowledges partial support as a Fellow in the Gutenberg Research College.




\begin{thebibliography}{99}

\bibitem{PDG}
  K.~Nakamura {\it et al.}  [Particle Data Group],
  J.\ Phys.\ G {\bf 37}, 075021 (2010) and 2011 partial update for the 2012 edition.

\bibitem{Czarnecki:2001pv} 
  A.~Czarnecki and W.~J.~Marciano,
  Phys.\ Rev.\ D {\bf 64}, 013014 (2001)
  [hep-ph/0102122].

\bibitem{DMzprime}
  P.~Fayet,
  Phys.\ Rev.\ D {\bf 70}, 023514 (2004)
  [hep-ph/0403226];
  D.~P.~Finkbeiner and N.~Weiner,
  Phys.\ Rev.\ D {\bf 76}, 083519 (2007)
  [astro-ph/0702587];
  N.~Arkani-Hamed, D.~P.~Finkbeiner, T.~R.~Slatyer and N.~Weiner,
  Phys.\ Rev.\ D {\bf 79}, 015014 (2009)
  [arXiv:0810.0713 [hep-ph]].

\bibitem{Davoudiasl:2010am} 
  A light $Z_d$ may also be invoked in asymmetric dark matter models to mediate efficient annihilation of the 
symmetric dark matter population.  See, {\it e.g.}, H.~Davoudiasl, D.~E.~Morrissey, K.~Sigurdson and S.~Tulin,
  Phys.\ Rev.\ Lett.\  {\bf 105}, 211304 (2010)
  [arXiv:1008.2399 [hep-ph]].

\bibitem{Holdom:1985ag} 
  B.~Holdom,
  Phys.\ Lett.\ B {\bf 166}, 196 (1986).

\bibitem{Davoudiasl:2012ag} 
  H.~Davoudiasl, H.~S.~Lee and W.~J.~Marciano,
  Phys.\ Rev.\ D {\bf 85}, 115019 (2012)
  [arXiv:1203.2947 [hep-ph]].
  
\bibitem{Fayet:2007ua} 
  P.~Fayet,
  Phys.\ Rev.\ D {\bf 75}, 115017 (2007)
  [hep-ph/0702176 [HEP-PH]];
  M.~Pospelov,
  Phys.\ Rev.\ D {\bf 80}, 095002 (2009)
  [arXiv:0811.1030 [hep-ph]].

\bibitem{Leveille:1977rc} 
  J.~P.~Leveille,
  Nucl.\ Phys.\ B {\bf 137}, 63 (1978).

\bibitem{McKeown:2011yj} 
  R.~D.~McKeown,
  AIP Conf.\ Proc.\  {\bf 1423}, 289 (2012)
  [arXiv:1109.4855 [hep-ex]].

\bibitem{Czarnecki:1900zz} 
  A.~Czarnecki and W.~J.~Marciano,
  in {\em Lepton Dipole Moments}, edited by L.~B.~Roberts and W.~J.~Marciano (World Scientific, Singapore, 2010), p11.

\bibitem{Hanneke:2008tm} 
  D.~Hanneke, S.~Fogwell and G.~Gabrielse,
  Phys.\ Rev.\ Lett.\  {\bf 100}, 120801 (2008)
  [arXiv:0801.1134 [physics.atom-ph]].
  
\bibitem{Bjorken:2009mm} 
  J.~D.~Bjorken, R.~Essig, P.~Schuster and N.~Toro,
  Phys.\ Rev.\ D {\bf 80}, 075018 (2009)
  [arXiv:0906.0580 [hep-ph]].

\bibitem{Abrahamyan:2011gv} 
  S.~Abrahamyan {\it et al.}  [APEX Collaboration],
  Phys.\ Rev.\ Lett.\  {\bf 107}, 191804 (2011)
  [arXiv:1108.2750 [hep-ex]].

\bibitem{Czarnecki:1995fw} 
  A.~Czarnecki and W.~J.~Marciano,
  Phys.\ Rev.\ D {\bf 53}, 1066 (1996)
  [hep-ph/9507420];
  A.~Czarnecki and W.~J.~Marciano,
  Int.\ J.\ Mod.\ Phys.\ A {\bf 15}, 2365 (2000)
  [hep-ph/0003049];
  A.~Czarnecki and W.~J.~Marciano,
  Nature {\bf 435}, 437 (2005).

\bibitem{Bennett:1999pd} 
  S.~C.~Bennett and C.~E.~Wieman,
  Phys.\ Rev.\ Lett.\  {\bf 82}, 2484 (1999)
  [Erratum-ibid.\ {\bf 82}, 4153 (1999)] [Erratum-ibid.\ {\bf 83}, 889 (1999)]
  [hep-ex/9903022];
  S.~L.~Gilbert, M.~C.~Noecker, R.~N.~Watts and C.~E.~Wieman,
  Phys.\ Rev.\ Lett.\  {\bf 55}, 2680 (1985).
  
\bibitem{Marciano:1990dp} 
  W.~J.~Marciano and J.~L.~Rosner,
  Phys.\ Rev.\ Lett.\  {\bf 65}, 2963 (1990)
  [Erratum-ibid.\  {\bf 68}, 898 (1992)].
    
\bibitem{Bouchiat:2004sp} 
  C.~Bouchiat and P.~Fayet,
  Phys.\ Lett.\ B {\bf 608}, 87 (2005)
  [hep-ph/0410260];
  C.~Bouchiat and C.~A.~Piketty,
  Phys.\ Lett.\ B {\bf 128}, 73 (1983).

\bibitem{Porsev:2009pr} 
  S.~G.~Porsev, K.~Beloy and A.~Derevianko,
  Phys.\ Rev.\ Lett.\  {\bf 102}, 181601 (2009)
  [arXiv:0902.0335 [hep-ph]];
  S.~G.~Porsev, K.~Beloy and A.~Derevianko,
  Phys.\ Rev.\ D {\bf 82}, 036008 (2010)
  [arXiv:1006.4193 [hep-ph]].

\bibitem{Derman:1979zc} 
  E.~Derman and W.~J.~Marciano,
  Annals Phys.\  {\bf 121}, 147 (1979).

\bibitem{Marciano:1982mm} 
  W.~J.~Marciano and A.~Sirlin,
  Phys.\ Rev.\ D {\bf 27}, 552 (1983);
  Phys.\ Rev.\ D {\bf 29}, 75 (1984)
  [Erratum-ibid.\ D {\bf 31}, 213 (1985)].

\bibitem{Dzuba:1985gw}
  V.~A.~Dzuba, V.~V.~Flambaum and I.~B.~Khriplovich,
  Z. Phys. D {\bf 1}, 243 (1986);
  E.~N.~Fortson, Y.~Pang and L.~Wilets,
  Phys.\ Rev.\ Lett.\  {\bf 65}, 2857 (1990);
  C.~Monroe, W.~Swann, H.~Robinson and C.~Wieman,
  Phys.\ Rev.\ Lett.\  {\bf 65}, 1571 (1990).

\bibitem{Anthony:2005pm} 
  P.~L.~Anthony {\it et al.}  [SLAC E158 Collaboration],
  Phys.\ Rev.\ Lett.\  {\bf 95}, 081601 (2005)
  [hep-ex/0504049].

\bibitem{E-08-016}
  JLab proposal E-08-016.
  
\bibitem{E-12-09-005}
  JLab proposal E-12-09-005.
  
\bibitem{MESA}
  K.~Aulenbacher,
  Hyperfine Interact.\ {\bf 200}, 3 (2011);
  H.~Spiesberger,
  to be published in DIS(2012) proceedings, Bonn, Germany.
  
\end{thebibliography}
\end{document}